\documentclass[]{article}%compile with pdflatex
\usepackage{graphicx}
\usepackage{amsmath}
\usepackage{authblk}
\usepackage{textcomp}\usepackage{gensymb}%\degree must load package textcomp before

\usepackage[numbers]{natbib}

\begin{document}

\title{Reinvestigation of the intrinsic magnetic properties of (Fe$_{1-x}$Co$_x$)$_2$B alloys and crystallization behavior of ribbons}% Force line breaks with \\
\author[1,2]{Tej~Nath Lamichhane}
\affil[1]{Department of Physics and Astronomy, Iowa State University, Ames, Iowa 50011, U.S.A.}
\affil[2]{Ames Laboratory, U.S. Department of Energy, Iowa State University, Ames, Iowa 50011, USA}
\author[2,4]{Olena Palasyuk}
\author[2]{Vladimir P. Antropov}
\author[3]{Ivan A. Zhuravlev}
\affil[3]{Department of Physics and Astronomy and Nebraska Center for Materials and Nanoscience, University of Nebraska-Lincoln, Lincoln, Nebraska 68588, USA}
\author[3]{Kirill D. Belashchenko}
\author[2]{Ikenna~C. Nlebedim}
\author[2]{Kevin~W. Dennis}
\author[1,2]{Anton Jesche}
\author[2,4]{Matthew~J. Kramer}
\affil[4]{Department of Materials Science and Engineering, Iowa State University, Ames, Iowa 50011, USA}
\author[1,2]{Sergey~L. Bud'ko}
\author[2]{R.~William McCallum}
\author[1,2]{Paul~C. Canfield}
\author[5,2]{Valentin Taufour}
%\ead{vtaufour@ucdavis.edu}
\affil[5]{Department of Physics and Astronomy, University of California Davis, Davis, California 95616, USA}
\maketitle
%\ead{vtaufour@ucdavis.edu}
%\address[5]{Department of Physics and Astronomy, University of California Davis, Davis, California 95616, USA}

%\nonumnote{for published version: https://doi.org/10.1016/j.jmmm.2020.167214}

\begin{abstract}
New determination of the magnetic anisotropy from single crystals of (Fe$_{1-x}$Co$_x$)$_2$B alloys are presented. The anomalous temperature dependence of the anisotropy constant is discussed using the standard Callen-Callen theory, which is shown to be insufficient to explain the experimental results. A more material specific study using first-principles calculations with disordered moments approach gives a much more consistent interpretation of the experimental data. Since the intrinsic properties of the alloys with $x=0.3-0.35$ are promising for permanent magnets applications, initial investigation of the extrinsic properties are described, in particular the crystallization of melt spun ribbons with Cu, Al, and Ti additions. Previous attempts at developing a significant hysteresis have been unsuccessful in this system. Our melt-spinning experiment indicates that this system shows rapid crystallization.
\end{abstract}

%\begin{keywords}
%temperature dependent anisotropy \sep single crystals \sep melt-spun ribbons
%\end{keywords}
\maketitle

\section{Introduction}

Magnetocrystalline anisotropy is one of the key ingredient for high-performance permanent magnets. In the rare-earth based magnets, the anisotropy comes mainly from the rare-earth $4f$ electrons. 
When, at sufficiently high temperatures, the anisotropy is uniaxial with a easy-axis for magnetization, large hysteresis and coercivity can be obtained resulting in high-performance permanent magnets. However, the criticality in the supply of rare-earth elements is driving researchers to look for new magnets with less or no rare-earth at all. One possible strategy is to take a closer look at the less studied $3d$ compounds~\cite{Kuzmin2014JPCM}.

Recently, we re-investigated the magnetic properties of (Fe$_{1-x}$Co$_x$)$_2$B alloys~\cite{Belashchenko2015APL}. The two end compounds Fe$_2$B and Co$_2$B have planar anisotropy at room temperature. However, a rather large uniaxial anisotropy is observed at some intermediate levels of substitutions with a maximum near $x=0.3$. Our new determination of the anisotropy constant $K_1$ at $2$\,K as a function of doping $x$ can be well reproduced by first-principles electronic structure analysis~\cite{Belashchenko2015APL}. The spin-reorientation can be understood by considering the filling of electronic bands with increasing electronic concentration~\cite{Belashchenko2015APL}. Another peculiar feature of (Fe$_{1-x}$Co$_x$)$_2$B alloys is that the change of anisotropy can also be observed by varying the temperature at a given composition~\cite{Iga1970JJAP415}. Since magnets are expected to operate at elevated temperatures, it is necessary to study the temperature dependence of the magneto-crystalline anisotropy. This is the purpose of this article, in~ which~ the~ temperature~ dependence of $K_1$ for (Fe$_{1-x}$Co$_x$)$_2$B alloys is reported. Measurements between $77$~K and $600$~K have already been reported and showed an anomalous dependence in which the magneto-crystalline anisotropy switches between planar and uniaxial anisotropy when the temperature is varied~\cite{Iga1970JJAP415}. Here, the measurements are extended both at lower temperatures (to $2$~K) and higher temperatures (to $1000$~K). Our results are in qualitative agreement with the previous report with a few quantitative differences. Our analysis shows that this anomalous temperature dependence cannot be fully described by the Callen and Callen (CC) theory~\cite{Callen1962JAP,Callen1966JPCS} which is the standard theory for the temperature dependence of the magneto-crys\-tal\-line anisotropy, while a first principle band structure analysis provides a rather satisfactory description of the magneto-crystalline anisotropy. In addition, magnetization, x-ray dif\-frac\-tion and differential scanning calorimetry measurements of melt-spun ribbons are presented. The results indicate that one of the major difficulty in developing coercivity in the (Fe$_{1-x}$Co$_x$)$_2$B alloys will be to control the rapid crystallization. The addition of Al can stop the crystallization but only temporarily, and crystallization occurs at $540$\degree C, or after a few months at room temperature.

\section{Methods}

Single crystals of (Fe$_{1-x}$Co$_x$)$_2$B were grown from high-temperature solution out of an excess of (Fe,Co) which was decanted in a centrifuge~\cite{CanfieldPMB1992}. The single crystals are grown as tetragonal rods which were cut using a wire saw to give them the shape of a rectangular prism. Magnetization measurements were performed in a Magnetic Property Measurement System (MPMS, Quantum Design) from $2$~K to $50$~K up to $5.5$~T and using a Vibrating Sample Magnetometer (VSM) in a Cryogen-free Physical Property Measurement System (Versalab, Quantum Design) from $50$~K to $350$~K up to $3$~T and up to $1000$~K using the oven option. An alumina cement (Zircar) was used to hold the sample on the heater stick for the high temperature measurements. The demagnetization factors along different directions were determined from the sample dimensions~\cite{Aharoni1998JAP}. The values along the easy magnetization direction were confirmed experimentally by using Arrott plots~\cite{Arrott1957PR}. The anisotropy constant $K_1$ was determined as the area between the two magnetization curves, with the field parallel and perpendicular to the $c$~axis, taken at the same temperature~\cite{Kuzmin2014JPCM}.

X-ray diffraction (XRD) was performed using a Rigaku Miniflex diffractometer (Cu-$K\alpha$ radiation). Lattice parameters were refined by the LeBail method using General Structure Analysis System (GSAS)~\cite{Larson2004} and EXPGUI~\cite{Toby2001JAC}. Instrument parameter files were determined from measurements on Si and Al$_2$O$_3$. XRD powder diffraction measurements on these standard materials have been performed regularly and allow to estimate the relative error for the given lattice parameters to be less than $0.002$~\cite{Jesche2014PM}. XRD for melt spun ribbons was carried out using a PANalytical X-Pert Pro Diffraction System (Co-$K\alpha$ radiation, $\lambda=1.78897$\,\AA).

Melt-spun ribbons of (Fe$_{1-x}$Co$_x$)$_2$B with $x=0.3$ and $0.35$ with 3wt\% Cu, 3wt\% Al, 4at.\% Ti, 8at.\% Ti were prepared. Melt spinning was performed in zero grade helium $1/3$\,atm. A $10$\,g ingot of arc-melted alloy was held in a SiO$_2$ crucible with a $0.8$\,mm orifice until superheated by induction to $150$\degree C above the liquidus, then ejected with $120$\,torr pressure onto a Oxygen-Free High Conductivity copper wheel rotating at $20$\,m/s for (Fe$_{0.7}$Co$_{0.3}$)$_2$B and $30$\,m/s for the other alloys. Differential scanning calorimetry (DSC) was performed using a simultaneous thermal analysis apparatus (STA) (Netzsch, DSC 404C). 
The STA measurements were performed in a helium atmosphere with a sweep rate of $20$\degree C/min up to $1300$\degree C.

For the electronic-structure calculations,~ we~ used the Green's function based linear muffin-tin orbital method with spin-orbit coupling included as a perturbation~\cite{Belashchenko2015APL}. Computations have been performed with $6$\,atoms per cell, a $24\times24\times24$ $k$-points grid for self-consistent calculation and a $30\times30\times30$ $k$-points grid for anisotropy calculations. Temperature has been included using the disordered local moment formalism with the computational details being similar to Refs.~\cite{Belashchenko2015APL,Zhuravlev2015PRL,Pujari2015PRL}, where the temperature dependence was studied earlier. Static density functional overestimation of local moment in Co$_2$B ($1.1$\,$\mu_B$ compared to experimental $0.76$\,$\mu_B$/Co) has been corrected by adding magnetic field using the procedure suggested in Ref.~\cite{Moriya1985,Antropov2011JAP}.

\section{Intrinsic Properties}

\begin{figure}[!htb]
\begin{center}
\includegraphics[width=7cm]{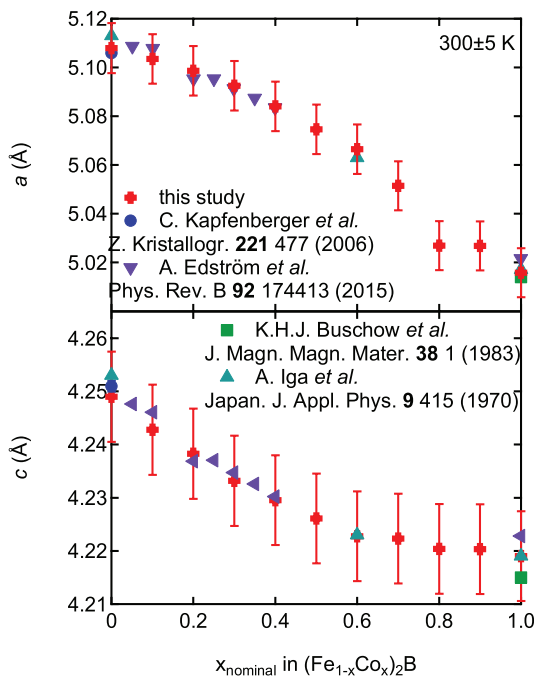}
\caption{\label{fig:Lattice}(Color online) Room temperature value of the lattice parameters as a function of $x$ in (Fe$_{1-x}$Co$_x$)$_2$B. Data from Refs.~\cite{Kapfenberger2006ZK,Buschow1983JMMM,Iga1970JJAP415,Edstrom2015PRB} are also reported.}
\end{center}
\end{figure}

Figure~\ref{fig:Lattice} shows the lattice parameters $a$ and $c$ as a function of $x$ in (Fe$_{1-x}$Co$_x$)$_2$B. Both $a$ and $c$ decrease monotonically with Co substitution, in agreement with previous reports~\cite{Iga1970JJAP415,Edstrom2015PRB}. The small deep observed in the $a$ axis at $x=0.8$ is smaller than the relative error, but interestingly, such an anomaly is expected from recent calculations in the generalized gradient approximation, treating disorder by the Virtual Crystal Approximation~\cite{Edstrom2015PRB}. In that theoretical study however, a small increase in the $c$ axis is also expected to occur at $x=0.8$ but is not observed in our experimental study.

\begin{figure}[!htb]
\begin{center}
\includegraphics[width=7cm]{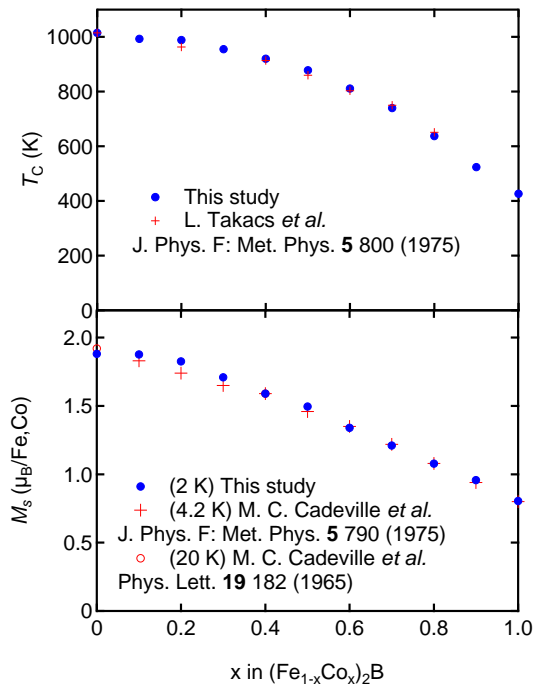}
\caption{\label{fig:TcandMs}(Color online) Curie temperature and spontaneous magnetization as a function of $x$ in (Fe$_{1-x}$Co$_x$)$_2$B. Data from Refs.~\cite{Cadeville1965PL,Cadeville1975JPFMP,Takacs1975JPFMP} are also reported.}
\end{center}
\end{figure}

The Curie temperature ~decreases ~monotonically ~from $1015$~K in Fe$_2$B to $426$~K in Co$_2$B as shown in Fig.~\ref{fig:TcandMs}. Similarly, Fig.~\ref{fig:TcandMs} shows that the spontaneous magnetization at low temperature decreases almost linearly from $1.9$~$\mu_B$/Fe in Fe$_2$B to $0.81$~$\mu_B$/Co in Co$_2$B in agreement with previous results~\cite{Cadeville1965PL,Cadeville1975JPFMP,Takacs1975JPFMP}. Calculations of the Curie temperature were recently performed with the density functional theory using the Coherent-Potential-Approximation to treat (Fe,Co) compositional disorder~\cite{Dane2015JPCM} and the results agree well with the experiments.

\begin{figure}[!htb]
\begin{center}
\includegraphics[width=7cm]{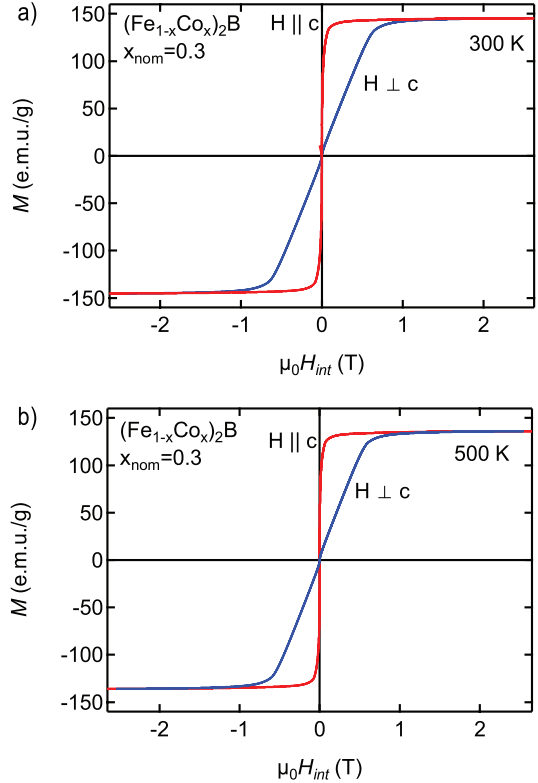}
\caption{\label{fig:MvsHab}(Color online) Magnetization versus internal magnetic field with the field applied parallel or perpendicular to the $c$-axis a) at $300$\,K, and b) at $500$\,K.}
\end{center}
\end{figure}

The field dependence of the magnetization for alloys of (Fe$_{1-x}$Co$_x$)$_2$B with $x=0.3$ is shown in Fig.~\ref{fig:MvsHab}. We can see the uniaxial anisotropy with the $c$-axis being the easy magnetization axis. In a tetragonal system, if we neglect the in-plane a\-ni\-so\-tro\-py, the anisotropy energy $E_A$ can be written~\cite{Millev1995PRB}:
\begin{eqnarray}
\dfrac{E_A}{V}&=&K_1\sin^2\theta+K_2\sin^4\theta
\end{eqnarray}
where $\theta$ is the polar angle of the magnetization direction. In (Fe$_{1-x}$Co$_x$)$_2$B, $K_2$ can be neglected~\cite{Iga1970JJAP415,Kuzmin2014JPCM} and the a\-ni\-so\-tro\-py is directly given by $K_1$ (we confirm this assumption later in this article). When $K_1$ is positive, the spontaneous magnetization is along the easy axis $c$-axis, whereas it is in the easy plane perpendicular to the $c$-axis when $K_1$ is negative.

\begin{figure*}[!htb]
\begin{center}
\includegraphics[width=14cm]{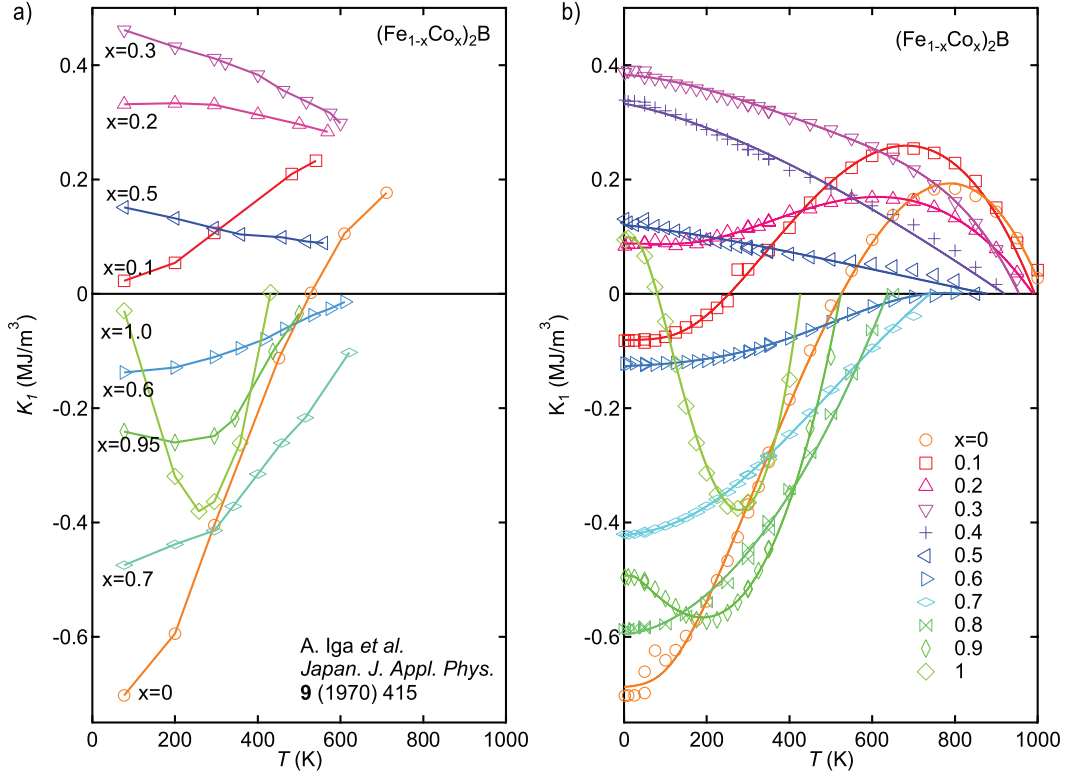}
\caption{\label{fig:K1vsT}(Color online) Temperature dependence of the anisotropy constant $K_1$ for various alloys of (Fe$_{1-x}$Co$_x$)$_2$B from the previous report~\cite{Iga1970JJAP415} in a) and from this study in b). Lines are guides to the eyes.}
\end{center}
\end{figure*}

The temperature dependence of the anisotropy constant $K_1$ for various alloys is reported in Fig.~\ref{fig:K1vsT}b. Our results are in qualitative agreement with the previous report~\cite{Iga1970JJAP415} (reproduced in Fig.~\ref{fig:K1vsT}a for comparison) and confirm the very anomalous temperature dependence of $K_1$ in this system. We note a few quantitative differences. For $x=0.1$, our results indicate a negative value of $K_1$ at low temperature whereas it ~seems ~to ~extrapolate ~to ~a ~positive value in Ref.~\cite{Iga1970JJAP415}. Interestingly, a negative value was also obtained in our recent density-functional calculations~\cite{Belashchenko2015APL}. Similarly, for $x=0.2$, a large value of $K_1\sim0.33$~MJ/m$^{3}$ was obtained in Ref.~\cite{Iga1970JJAP415}, whereas our results indicate $K_1\sim0.08$~MJ/m$^{3}$, in~~better ~~agreement ~with ~~the ~~theoretical ~value ~of $\sim0.07$~MJ/m$^{3}$~\cite{Belashchenko2015APL}. For $x=1$, i.e. for Co$_2$B, our measurements indicate that the anisotropy is axial ($K_1>0$) at lower temperatures whereas it becomes planar ($K_1<0$) at higher temperatures. A positive value of $K_1$ was expected from extrapolation of the previous results to lower temperatures~\cite{Iga1966JPSJ,Iga1970JJAP415} and a corresponding anomaly was observed near $70$~K in ac susceptibility measurements~\cite{Bormio2010JAC}. We note that this spin reorientation is suppressed with Fe substitution since it is not observed for $x=0.9$ in our data and for $x=0.95$ in Ref.~\cite{Iga1970JJAP415}. Using M\"ossbauer spectroscopy, it was found that the reorientation occurs between $1.6$ and $4.2$~K with $1$~at.\%~$^{57}$Fe~\cite{Morimoto1988HI}. Recent density-functional-theory calculations~\cite{Belashchenko2015APL,Edstrom2015PRB} were not able to reproduce the positive sign of the magnetocrystalline anisotropy energy of Co$_2$B. However, good agreement is obtained on the Fe-rich side, in particular for the optimal value at $x=0.3$.

We now comment on the anomalous temperature dependence of $K_1$ in this system. In principle, the temperature dependence of the anisotropy constant is given by the so-called Callen-and-Callen (CC) law~\cite{Callen1960JPCS,Callen1962JAP,Callen1963PR,Callen1966JPCS}. In a system with uniaxial anisotropy, the temperature dependence of the anisotropy constants are given by:
\begingroup\makeatletter\def\f@size{9}\check@mathfonts
\def\maketag@@@#1{\hbox{\m@th\large\normalfont#1}}%
\begin{eqnarray}
K_1(T)&=&\left(K_1^0+\dfrac{7}{8}K_2^0\right)\left(\dfrac{M(T)}{M(0)}\right)^3-\dfrac{7}{8}K_2^0\left(\dfrac{M(T)}{M(0)}\right)^{10}~~~~~~\label{eqn:K1full}\\
K_2(T)&=&K_2^0\left(\dfrac{M(T)}{M(0)}\right)^{10}
\end{eqnarray}\endgroup
Since $K_2^0$ is usually negligible, equation~\ref{eqn:K1full} is often reduced to:
\begin{eqnarray}
K_1(T)&=&K_1^0\left(\dfrac{M(T)}{M(0)}\right)^3\label{eqn:K1simple}
\end{eqnarray}
It follows from equation~\ref{eqn:K1simple} that the temperature dependence of $K_1$ is monotonic and cannot reproduce the observed change of sign. However, large values of $K_2^0$ can induce a change of sign in the temperature dependence of $K_1(T)$~\cite{Millev1995PRB}. Therefore, before claiming that the CC law is violated, it is necessary to confirm the assumption that $K_2^0$ is negligible.

\begin{figure}[!htb]
\begin{center}
\includegraphics[width=8.3cm]{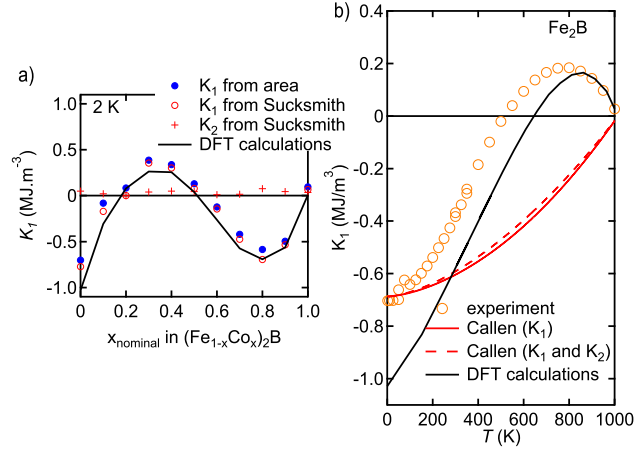}
\caption{\label{fig:Callenab}(Color online) a) Comparison of the anisotropy energy constants $K_1$ and $K_2$ determined by two techniques: area between the two magnetization curves, with the field parallel and perpendicular to the $c$~axis, taken at $2$~K (neglecting $K_2$); or using the Sucksmith method~\cite{Sucksmith1954PRSLA,Neel1960JAP} ($K_1$ and $K_2$). b) Temperature dependence of $K_1$ for Fe$_2$B. The expected behavior from the CC law is shown in the cases when $K_2^0$ is neglected (full line, eqn.~\ref{eqn:K1simple}) or not (dashed line, eqn.~\ref{eqn:K1full}). Density functional calculations are shown by black lines.}
\end{center}
\end{figure}

In order to determine the anisotropy constant $K_2$, we used the Sucksmith method~\cite{Sucksmith1954PRSLA,Neel1960JAP}. In this method, the first and second order uniaxial anisotropy constants, $K_1$ and $K_2$, can be determined by plotting $H/M$ vs $M^2$ and by fitting the linear part to:
\begin{eqnarray}
\dfrac{\mu_0H}{M_\bot}=&\dfrac{2K_1}{{M_s}^2}+\dfrac{4K_2}{{M_s}^4}{M_\bot}^2~~~~~~~~~~~~&\textrm{(easy axis)~~~~~~}\\%\nonumber\\
\dfrac{\mu_0H}{M_{||}}=&-\dfrac{2K_1+4K_2}{{M_s}^2}+\dfrac{4K_2}{{M_s}^4}{M_{||}}^2~~~&\textrm{(easy plane)~~~~~~}%\nonumber
\end{eqnarray}
where $M_\bot$ ($M_{||}$) is the magnetization measured perpendicular (parallel) to the easy (hard) axis. $M_s$ can be obtained from the magnetization curves along the easy axis. The obtained values of $K_1$ and $K_2$ are shown in Fig.~\ref{fig:Callenab}a. We can see that $K_2$ is indeed negligible. The effect of $K_2$ on the expected temperature dependence of $K_1$ in the CC theory is illustrated as a dashed line in Fig.~\ref{fig:Callenab}b where we can see that considering $K_2$ produces only a small correction. It can therefore be concluded that $K_1(T)$ in (Fe$_{1-x}$Co$_x$)$_2$B does not follow the CC law.

A modification of the CC law can be made to account for the effect of thermal expansion~\cite{Carr1958PR}. Due to anisotropic thermal expansion, $K_1^0$ has a temperature dependence~\cite{Brenner1957PR} so that $K_1(T)$ is written in the form~\cite{Carr1958PR}:
\begin{eqnarray}
K_1(T)&=&K_1(0)(1-u\alpha_{c/a} T)\left(\dfrac{M(T)}{M(0)}\right)^3
\end{eqnarray}
where $\alpha_{c/a}$ is an average thermal expansion coefficient for $c/a$ and $u$ is a constant determined experimentally.
Such formula was used to describe the temperature dependence of $K_1$ in Co$_2$B~\cite{Iga1966JPSJ}. However, the thermal expansion of $c/a$ is very small in (Fe$_{1-x}$Co$_x$)$_2$B~\cite{Iga1970JJAP415} and it is unlikely that it could be responsible for a change of sign of the anisotropy. 

The established above failure of the CC model to describe $K_1(T)$~ is~ not~ very~ surprising~ in~ a~ system~ such as (Fe$_{1-x}$Co$_x$)$_2$B alloys, which are metallic alloys with magnetic moments being itinerant to a large extent. The magnetic anisotropy is not expected to have a pure single ion origin and there are several atomic components with magnetic states strongly depending on the chemical composition. %A huge deviation from CC model axiomatics is the fact that magnetic anisotropy does not follow magnetization as temperature raises.n 
All these facts possibly lead to the clear deviation from the CC model predictions, and serve as a clear indication of the presence of a very different and more complicated physical mechanism for the magnetic anisotropy.

Alternatively, one can study the magnetic anisotropy energy (MAE) using ab-initio electronic structure analysis of these alloys. In this approach, no assumptions related to the MAE structure are made and a possible itineracy of the conduction electrons in metals is taken into account.

Two major effects contribute to the unusual concentration dependence of the anisotropy: a change of the exchange splitting and a band broadening. Both spin-conserved and spin-flip transitions contribute significantly to the anisotropy. Each spin contribution depends strongly on the position of the minority spin bands. For instance, while at $T=0$\,K the minority spins dominate and stabilize the magnetic moments in plane (negative sign), at the Curie temperature the majority spins already reorient their moments along the $z$-direction (positive sign). Thus, the spin reorientation is determined by this spin polarization change of dominating electrons. In addition, due to such mechanisms, in a certain range of the amplitudes of magnetic moments, the resulting anisotropy can increase with decreasing magnetization. Such band structure effect can be common in metallic itinerant magnets and cannot be described by such simple single ion anisotropy model such as the CC model.

\begin{figure}[!htb]
\begin{center}
\includegraphics[width=6.5cm]{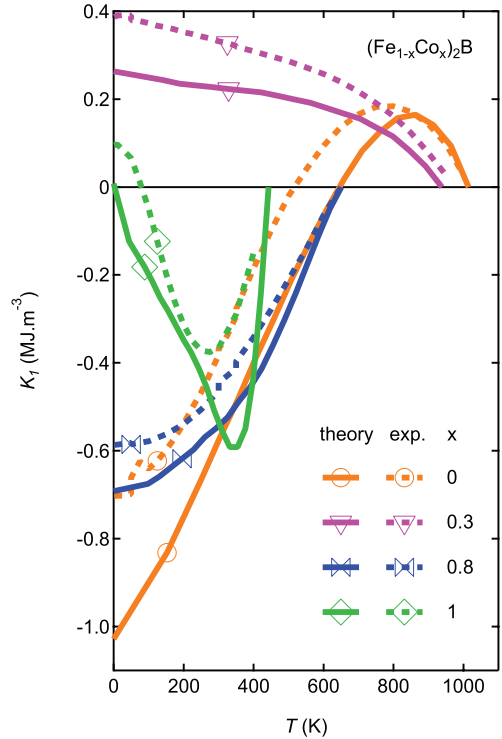}
\caption{\label{fig:exp_theor_v2}(Color online) A comparison of the experimental and theoretical temperature dependencies of the anisotropy constant $K_1$ for various alloys of (Fe$_{1-x}$Co$_x$)$_2$B.}
\end{center}
\end{figure}

\begin{figure}[!htb]
\begin{center}
\includegraphics[width=6.5cm]{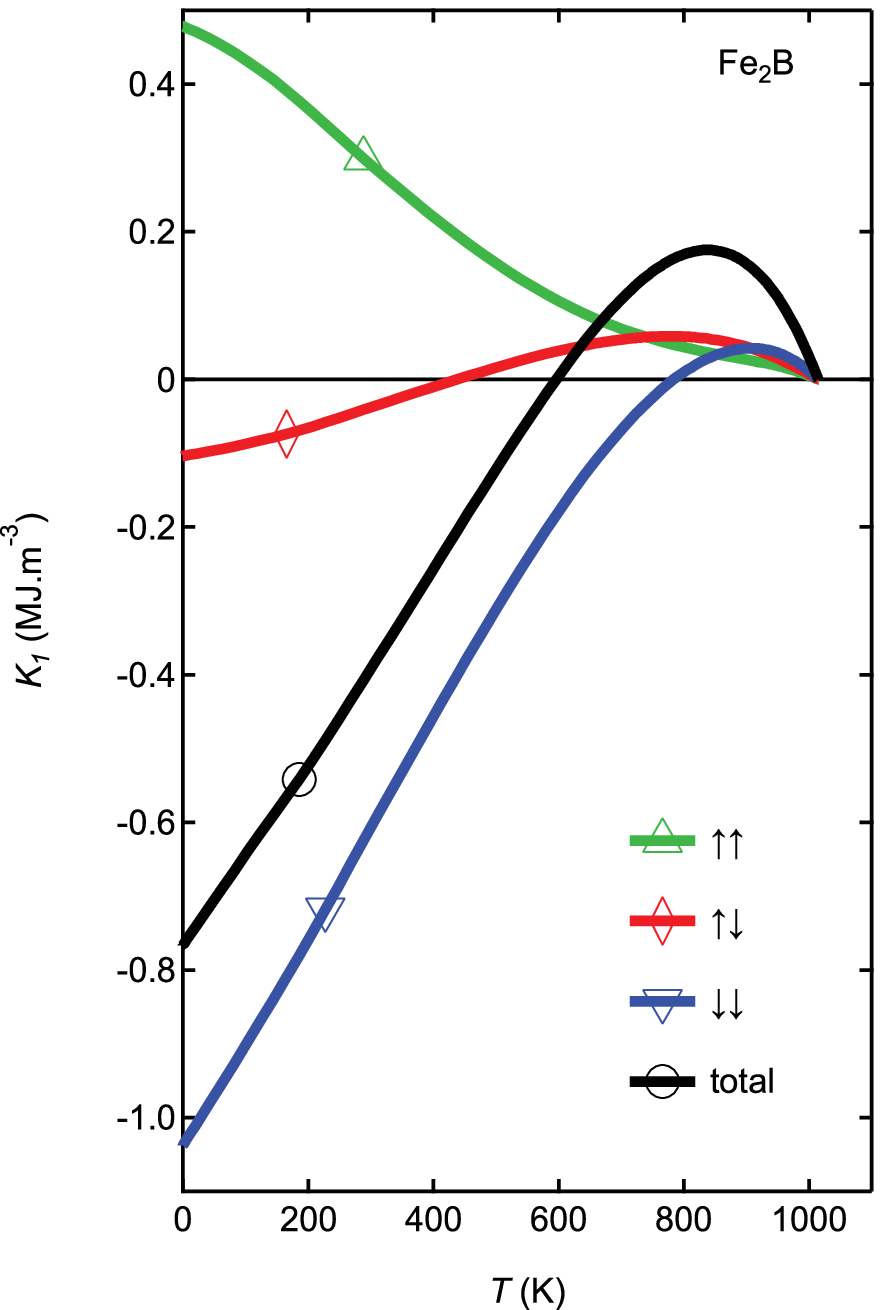}
\caption{\label{fig:Fe2B_pure_Kso_temp_dep_22}(Color online) Total magnetic anisotropy $K_{SO}/2$ for Fe$_2$B  (black line) and its spin components as a function of temperature}
\end{center}
\end{figure}

To demonstrate how different scattering processes contribute to the anisotropy, we decompose the spin-orbit coupling anisotropy, following the prescription from Ref.~\cite{Antropov2014SSC}. In this case, the total anisotropy can be presented as $K=K_{SO}/2$, where the spin-orbit coupling anisotropy $K_{SO}$ in turn can be decomposed into different spin channels contributions $K_{\sigma\sigma'}$. In Fig.~\ref{fig:Fe2B_pure_Kso_temp_dep_22}, we show $K_{SO}/2$ for Fe$_2$B and its spin components. Clearly, different spin channels have contributions with different signs, amplitudes and very different dependencies on temperature reflecting the complicated character of the magnetic anisotropy in metallic systems. 
The strong concentration dependence is due to the modification of the character of electronic bands near the Fermi level with chemical doping. All these results demonstrate a rich physics of anisotropic phenomena in metals.

\section{Extrinsic Properties: crystallization of melt-spun ribbons}

Although large single crystals with minimal defects, are ideal to investigate the intrinsic properties, and in particular the magnetic anisotropy, they cannot be used as permanent magnets. This is because magnetic domains can form easily in the absence of pinning by defects and almost no coercivity can be obtained. In order to develop coercivity, it is necessary to control the defects, and the macro and microstructure. We now turn to the investigation of the extrinsic properties in (Fe$_{1-x}$Co$_x$)$_2$B alloys. Since the magnetic anisotropy is axial and the highest at $x=0.3-0.35$, we focus our efforts near that composition. In order to provide pinning sites for the magnetic domains, a possibility is to add other phases to the composition. We note that a coercivity of $30$\,kA/m ($380$\,Oe) was recently observed on melt spun ribbons with a nominal composition of (Fe$_{0.7}$Co$_{0.3}$)$_{71}$B$_{29}$, i.e. between (Fe$_{1-x}$Co$_x$)$_2$B and (Fe$_{1-x}$Co$_x$)$_3$B~\cite{Wallisch2015JAC}. In heavily milled (Fe$_{0.675}$Co$_{0.3}$Re$_{0.025}$)$_2$B with and without excess boron, coercivity near $900$\,Oe was obtained after annealing~\cite{Kim2018IEEETM}. In this study, we attempted to control the crystallization by producing melt-spun ribbons with the addition of Al, Cu, and Ti which are known to improve magnetic properties in other Fe or Co based magnets~\cite{Liu2005,Mottram2000JMMM,Cong2012JN,Hono1999AM,Kadin1986MRSP}.

\begin{figure}[!htb]
\begin{center}
\includegraphics[width=7cm]{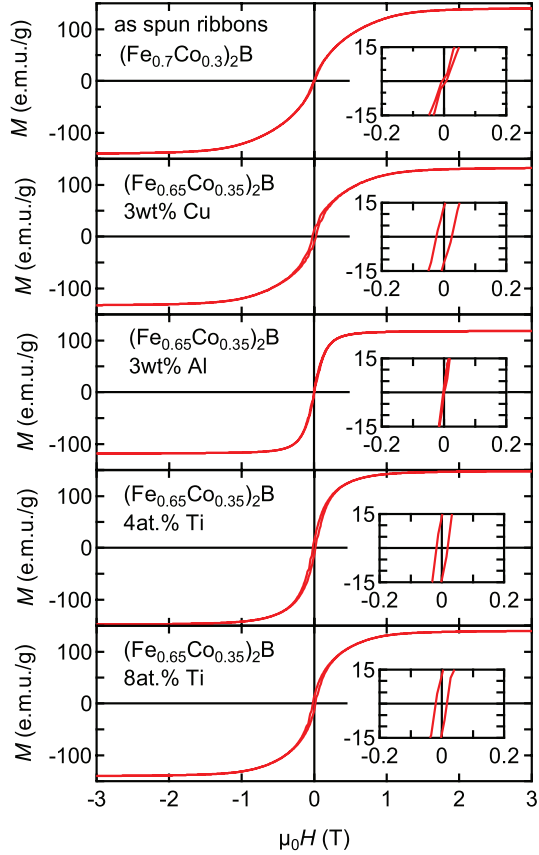}
\caption{\label{fig:MvsHasspun}(Color online) Magnetic hysteresis loop of as spun ribbons of (Fe$_{1-x}$Co$_x$)$_2$B with $x=0.3$ and $0.35$, with 3wt\% Cu, 3wt\% Al, 4wt\% Ti, 8wt\% Ti. The insets are zoom in the low field region.}
\end{center}
\end{figure}

The magnetic hysteresis loop of as spun ribbons with wheel speed of $20$ or $30$~m/s are shown in Fig.~\ref{fig:MvsHasspun}. No detectable hysteresis can be observed for the pure (Fe$_{1-x}$Co$_x$)$_2$B alloys as well as with $3$ wt\% Al. The alloys with $3$ wt\% Cu have a coercive field of $220$\,Oe and the alloys with 4 and 8\,at.\% Ti both have a coercive field of $160$\,Oe. These small values confirm the difficulty to develop hysteresis in these alloys~\cite{Wallisch2015JAC,Kim2018IEEETM}. In order to understand the small values of coercivity, we performed x-ray powder diffraction and DSC analysis.

\begin{figure}[!htb]
\begin{center}
\includegraphics[width=7cm]{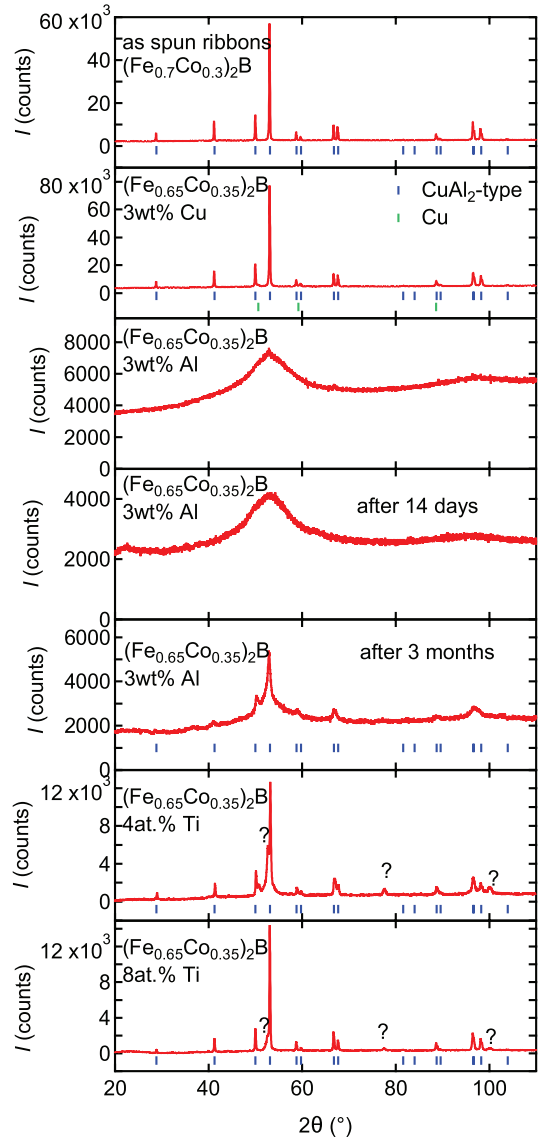}
\caption{\label{fig:xrayasspun}(Color online) Powder x-ray diffraction of the as spun (Fe$_{1-x}$Co$_x$)$_2$B ribbons with $x=0.3$ and $0.35$, with 3wt\% Cu, 3wt\% Al, 4wt\% Ti, 8wt\% Ti, as well as the 3wt\% Al ribbons after 14 days and 3 months.}
\end{center}
\end{figure}

As ~can ~be ~seen ~on figure~\ref{fig:xrayasspun}, the ~as ~spun ~ribbons ~of the (Fe$_{1-x}$Co$_x$)$_2$B alloys (pure, or with Cu and Ti additions) are already crystalline, indicating a rapid crystallization of this system upon cooling. We note that additional diffraction peaks corresponding to Cu are observed in the (Fe$_{0.65}$Co$_{0.35}$)$_2$B ribbons with 3wt\% Cu, and few additional peaks of unidentified phases are also observed in the ribbons with Ti additions. The presence of impurity phases can explain the origin of the small coercivity observed in these alloys. On the other hand, the addition of 3 wt\% Al, is successful in retarding the crystallization. However, we can see that the crystallization occurs after $3$\,months at ambient temperature.

\begin{figure}[!htb]
\begin{center}
\includegraphics[width=7cm]{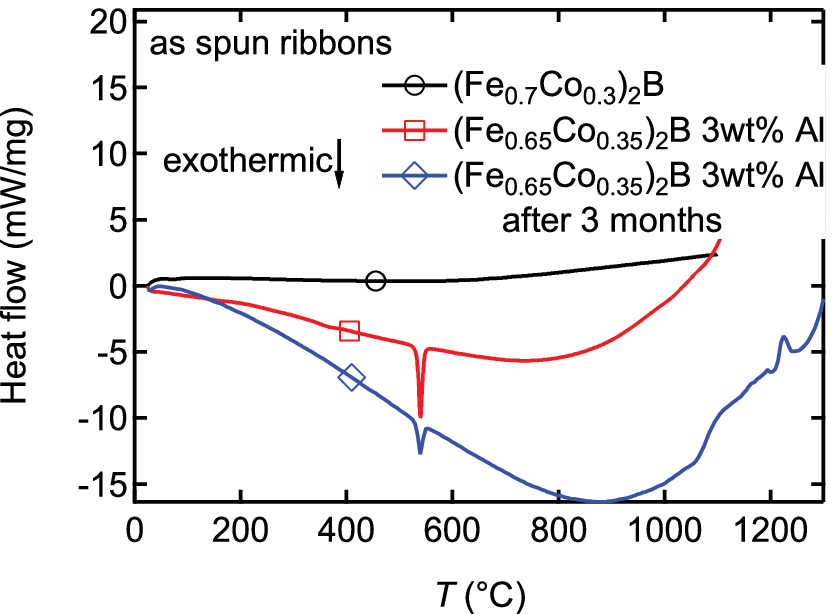}
\caption{\label{fig:DSC}(Color online) DSC plot during heating of the (Fe$_{1-x}$Co$_x$)$_2$B ribbons with $x=0.3$ (as spun) and $x=0.35$ with 3wt\% Al (as spun and after 3 months).}
\end{center}
\end{figure}

Figure~\ref{fig:DSC} shows the DSC data for (Fe$_{1-x}$Co$_x$)$_2$B ribbons with $x=0.3$ (as spun) and $x=0.35$ with 3wt\% Al (as spun and after 3 months).
There is no crystallization peak for the (Fe$_{0.7}$Co$_{0.3}$)$_2$B in agreement with the fact that the powder x-ray diffraction indicates that (Fe$_{1-x}$Co$_x$)$_2$B has already crystallized during the melt-spinning. On the other hand, there is a clear crystallization peak at $540$\degree C for the alloy with 3wt\% Al. The peak is smaller (lower area under the curve) for alloys that were annealed at room temperature for 3 months, in agreement with the partial crystallization in the powder x-ray diffraction pattern.

A desired approach would be that Al additions promote amorphization during the synthesis, such that grains can grow uniformly during crystallization. Our results show that Al additions combined with melt-spinning with an injection temperature of $1450$ \degree C and wheel-speed of $30$\,m/s will not be sufficient to control the crystallization on a time-scale long enough for permanent magnet applications, and that other methods to control the macro and microstructure will be necessary. One of the major difficulty in developing coercivity in the (Fe$_{1-x}$Co$_x$)$_2$B alloys will be to control the rapid crystallization. The addition of Al can stop the crystallization but only temporarily, and crystallization occurs at $540$\degree C, or after a few months at room temperature.

\section{Conclusions}

In conclusion, we have re-investigated the intrinsic and extrinsic magnetic properties of (Fe$_{1-x}$Co$_x$)$_2$B alloys. The temperature dependence of the anisotropy constant $K_1$ is largely anomalous with some temperature induced changes of sign which cannot be accounted for by the Callen-and-Callen model. Instead, our realistic electronic structure analysis produces a remarkably successful description of the temperature and concentration dependence of the anisotropy in these metallic itinerant magnets. The alloys with $x=0.3-0.5$ are the most promising for permanent magnet applications. However, previous attempts at developing a significant hysteresis have been unsuccessful in this system. Our melt-spinning experiment indicates that this system shows rapid crystallization. Further studies will be necessary to control the crystallization and develop the extrinsic properties in these materials.

\section*{Acknowledgements}

We would like to thank D.~Finnemore, M. D\"ane for useful discussions.
The research was supported by the Critical Materials Institute, an Energy Innovation Hub funded by the U.S. Department of Energy, Office of Energy Efficiency and Renewable Energy, Advanced Manufacturing Office and by the Office of Basic Energy Science, Division of Materials Science and Engineering. The work at UNL was supported by the National Science Foundation through Grant No. DMR-1609776 and performed utilizing the Holland Computing Center of the University of Nebraska. Ames Laboratory is operated for DOE by Iowa State University under contract No. DE-AC02-07CH11358.

\end{document}